# Core-shell droplets and microcapsules formed through liquid-liquid phase separation of a colloid-polymer mixture

Steven Dang,[a] John Brady,[a] Ryle Rel,[a] Sreenidhi Surineni,[a] Conor O'Shaughnessy,[a] and Ryan McGorty *,[a]

[a]Department of Physics and Biophysics, University of San Diego, San Diego, CA 92110
*E-mail: rmcgorty@sandiego.edu

Microcapsules allow for the controlled containment, transport, and release of cargoes ranging from pharmaceuticals to fragrances. Given the interest from a variety of industries in microcapsules and other core-shell structures, a multitude of fabrication strategies exist. Here, we report on a method relying on a mixture of temperature-responsive microgel particles, poly(N-isopropylacrylamide) (pNIPAM), and a polymer which undergo fluid-fluid phase separation. At room temperature this mixture separates into colloid-rich (liquid) and colloid-poor (gas) fluids. By heating the sample above a critical temperature where the microgel particles shrink dramatically and develop a more deeply attractive interparticle potential, the droplets of the colloid-rich phase become gel-like. As the temperature is lowered back to room temperature, these droplets of gelled colloidal particles reliquefy and phase separation within the droplet occurs. This phase separation leads to colloid-poor droplets within the colloid-rich droplets surrounded by a continuous colloid-poor phase. The gas/liquid/gas all-aqueous double emulsion lasts only a few minutes before a majority of the inner droplets escape. However, the colloid-rich shell of the core-shell droplets can solidify with the addition of salt. That this method creates a core-shell structures with a shell composed of stimuli-sensitive microgel colloidal particles using only aqueous components makes it attractive for encapsulating biological materials and making capsules that respond to changes in, for example, temperature, salt concentration, or pH.

## 1 Introduction

Encapsulation of material in droplets or microcapsules has a range of applications from drug delivery to the controlled release of fragrances and flavors[1-6]. In such applications, materials within the microcapsule must be separated from the outer environment by a solid or liquid shell. Various methods for microcapsule fabrication exist, too numerous to review here[7,8]. But we highlight a couple methods that share similarities with what we present. First, colloidal particles can be used to create a semi-permeable solid shell[9,10]. A common strategy for colloidal capsule synthesis is to start with a Pickering emulsion, where fluid droplets are stabilized against coalescence with colloidal particles adsorbed to the interface, and have the colloidal particles lock in place and form a rigid shell through, for example, thermal annealing[11] or cross-linking[12], producing what are termed colloidosomes[13]. Second, double emulsions may be used for compartmentalization or as templates for forming solid-shell capsules[14,15]. A double emulsion is formed when core droplets reside within larger droplets surrounded by a continuous phase[16-18]. The fluid shell separating the inner core droplet from the outer continuous phase can be solidified using, for example, photo-sensitive agents and controlled exposure to light[19-22] or having a solvent in the shell evaporate thereby concentrating what remains[23,24]. Sharing aspects of both broad categories of microcapsule formation, in the method we present here a shell of colloidal particles is formed after a colloid-polymer mixture which phase separates into two fluid phases is made to form a colloid-poor/colloid-rich/colloid-poor double emulsion.

In recent years, all aqueous formulations of multiple emulsions and microcapsules have been devised. Spurred by a desire to avoid organic solvents and enhance biocompatibility as well as to make protocells or *in vitro* analogs of intracellular membraneless organelles[25,26], these methods often use mixtures of two or more water-soluble polymers that phase separate into distinct fluid phases[27-30]. Synthesis of microcapsules using all aqueous components is challenged by the typically low surface tension (often ~ µN/m). However, both double emulsions and colloidosome-like structures have been formed using all aqueous components[31-37]. Our described method also results in a shell of colloidal particles using all-aqueous components, but rather than having colloids assemble at the interface between two aqueous fluids, we have a colloid-rich fluid as one of our aqueous phases.

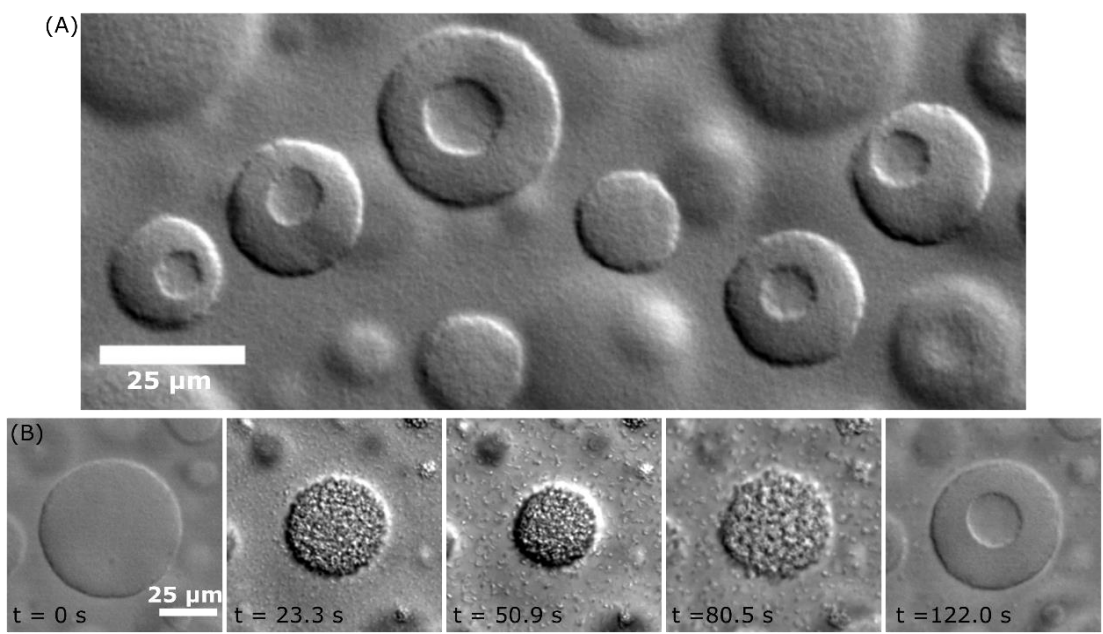

Fig. 1 Core-shell droplets form from a colloid-polymer mixture. (A) Multiple core-shell droplets with an inner colloid-poor phase, a shell of a colloid-rich phase, and a continuous colloid-poor phase are imaged with a 20× objective using DIC. (B) The process of core-shell droplets is shown in this time sequence of images. At room temperature at t = 0 s, we observe droplets of the colloid-rich phase. The sample is heated with a heat gun starting at about t = 16 s for approximately 20 s. After heating, the pNIPAM particles within the droplet are in the shrunken state and have formed a solid-like aggregate (t = 22.3 s and t = 50.9 s). As the sample cools, we observe the gelled pNIPAM particles reliquefy (t = 80.5 s) and the droplet contains numerous colloid-poor droplets inside which coalesce into a single inner droplet (t = 122 s). A movie of this process is shown in the Supplemental Materials (Movie S1).

In this paper, we report on how a temperature-responsive colloid-polymer mixture can lead to double emulsions and microcapsules. We use poly(N-isopropylacrylamide) (pNIPAM) microgel particles which exhibit a sharp change in size at a critical temperature of ~33 °C, termed the lower critical solution temperature (LCST). Through the addition of the polymer xanthan, depletion attraction between the pNIPAM particles leads to fluid-fluid phase separation at room temperature. Over the range of concentrations we explore, mixing and manually shaking samples leads to an emulsion of colloid-rich droplets that are ~10s of microns in diameter surrounded by a colloid-poor continuous phase. Raising the temperature of the sample above pNIPAM's LCST causes the colloid-rich droplets to form a solid gel due to the increased particle-particle attraction[38-41]. As the sample cools back to room temperature, we observe fluid-fluid phase separation within the colloid-rich droplet. This leads to a colloid-poor phase inner droplet within a colloid-rich phase droplet surrounded by a continuous colloid-poor phase—a variety of water-in-water-in-water (W/W/W) double emulsion (see Fig. 1). To create long-lived microcapsules with a solid shell, we solidify the colloid-rich phase by increasing the salt concentration.

## 2 Experimental

### 2.1 pNIPAM microgel particle synthesis

We synthesized pNIPAM microgel particles guided by previously published protocols[42]. We use ammonium persulfate (APS), N-isopropylacrylamide (NIPAM), N,N′- methylenebisacrylamide (BIS), and sodium dodecyl sulfate (SDS), all used as received without purification. Synthesis was done in a 250 mL three necked round bottom flask heated to 80 °C and under nitrogen atmosphere. In 200 mL of deionised water, we added 3.46 g of NIPAM, 0.15 g of BIS and 0.11 g of SDS. Once dissolved, this solution was transferred to the three necked flask. We then added 0.11 g of APS dissolved in 3 mL of deionised water. We let the reaction run for 4 hours with constant stirring with a magnetic stir bar and continuous bubbling of nitrogen through the solution. After synthesis, we dialyzed the solution against deionized water for two weeks.

### 2.2 Characterization of pNIPAM particles

To estimate the pNIPAM particle volume fraction, we measured the viscosity of suspensions of the particles in deionized water using a Cannon-Ubbelohde dilution viscometer. According to the Einstein-Bachelor relation the ratio of the viscosity of a suspension of spherical particles to that of the suspending medium is related to the particle volume fraction as $\eta/\eta_0 = 1 + 2.5\phi$ where $\phi$ is the effective particle volume fraction[43]. We found the volume fraction of our stock solution of pNIPAM particles to be $\phi_p = 0.38$.

We found the temperature-dependent size of our pNIPAM particles using dynamic light scattering measurements taken with a Malvern Zetasizer NanoZS. Over a temperature range of 23 to 40 °C we measured the size every 0.5 °C.

### 2.3 Preparation of pNIPAM-xanthan samples

We use xanthan polymer, Ticaxan® Xanthan VI from TIC Gums, to act as a depletant. We note that xanthan has been used in previous colloid-polymer samples to generate fluid-fluid phase separation[44, 45]. As reported by the manufacturer, the molecular weight is in the range of 4 to 12 × 10$^6$ g mol$^{-1}$. We prepare a stock solution of xanthan with a concentration of 0.2% by weight (wt%) with 0.1 M NaCl and 2 mM NaN$_3$.

### 2.4 Microscopy

All microscopy data was acquired using a Nikon Eclipse Ti inverted microscope with either a 20× Nikon Plan Apo λ (0.75 NA) or a 60× oil lens (1.40 NA). We used differential interference contrast (DIC) to enhance the contrast between the colloid-rich and colloid-poor phases. Images were captured with an sCMOS camera, pco.panda 4.2.

Samples imaged on the microscope were loaded into a chamber made from a glass slide and coverslip. The chambers were approximately 100 to 200 μm thick, held ~20 μL of sample, and were sealed with epoxy.

### 2.5 Heating experiments

To heat the samples, atop our Nikon inverted microscope we use a stage top incubator from Tokai Hit (model INU-WSKM) or a heat gun (Genesis, 750W). When using the Tokai Hit stage top incubator, we did not employ an objective lens heater. Therefore, the temperature reported by a probe placed ~2 cm from the sample is likely higher than the temperature of the region of the sample being imaged. For experiments with the heat gun, we aimed the gun at the sample on the microscope ~20 cm away and turned the gun on for ~10-20 seconds.

### 2.6 Rheology

Rheological measurements were done using a Discovery Hybrid Rheometer from TA Instruments. The sample was loaded between 40-mm-diameter stainless steel parallel plates with a gap of 140 μm. To find the storage and loss moduli as a function of temperature we performed frequency sweeps from 100 rad/s to 0.1 rad/s with 3.0% strain amplitude every 1 °C from 20 to 36 °C. The sample was given 3 minutes to equilibrate at each temperature.

To measure the rheology of just the colloid-rich phase we made a 1.0 mL mixture of pNIPAM and xanthan (0.08 wt% xanthan, $\phi_p$ = 11.4%). The mixture sat for approximately 60 hours to allow the colloid-rich phase to sediment. We pipetted off the colloid-poor phase and used only the colloid-rich phase for rheological measurements.

## 3 Results and discussion

An aqueous suspension of pNIPAM colloidal particles and solution of xanthan are mixed at various ratios, vigorously shaken by hand, and observed under an optical microscope. For certain concentrations of colloid and polymer we observe fluid-fluid phase separation evidenced by colloid-rich droplets surrounded by a colloid-poor continuous phase. Given the size and refractive index of these pNIPAM particles, we cannot resolve individual particles, but we can clearly identify the interface between the two fluid phases, especially using differential interference contrast (DIC) which is the modality used in all micrographs shown. While the density difference between the colloid-rich and colloid-poor phases leads to sedimentation, that density difference is low enough to permit the observation of droplets in our sample contained in a glass slide of ~100-200 μm thick for at least an hour.

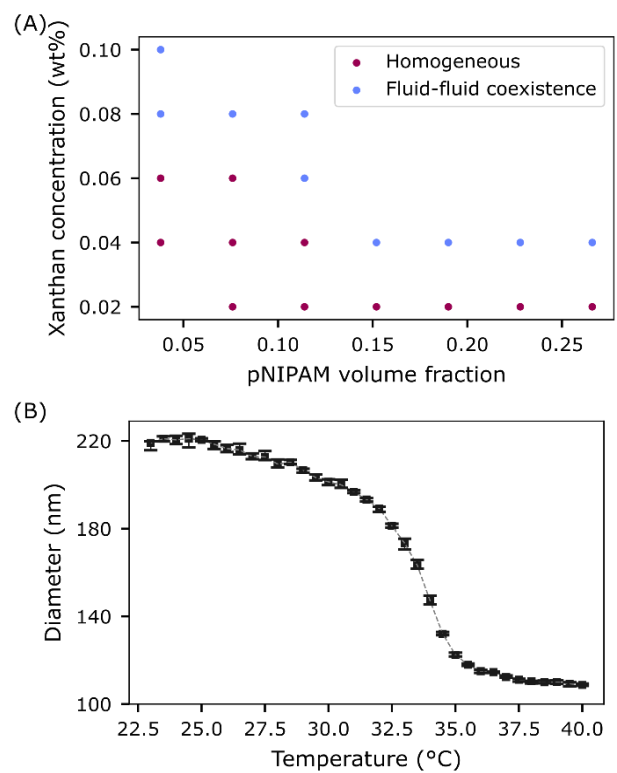

Fig. 2 Our pNIPAM-xanthan mixture exhibits fluid-fluid phase separation at certain concentrations and temperatures. (A) The phase diagram of pNIPAM-xanthan mixtures at room temperature is shown with the red points indicating concentrations for which the sample is mixed and blue indicating concentrations for which the sample has demixed into two fluid phases. (B) Since the pNIPAM particles shrink as the temperature increases, the phase behavior of our mixture is temperature dependent. Dynamic light scattering reveals the change in diameter of our pNIPAM particles. Particles at room temperature have a size of ~220 nm. The particles shrink dramatically around the LCST of approximately 33 °C.

By inspecting pNIPAM-xanthan samples under a microscope, typically with a 20× objective, we create the phase diagram shown in Fig. 2A. This phase diagram was constructed with all samples at room temperature and will not be the case for other temperatures due to the temperature-responsive property of pNIPAM. As the temperature increases, the diameter of pNIPAM particles decreases from about 220 nm at room temperature to about 110 nm at 40 °C, as shown in Fig. 2B. The dramatic decrease of the particles' size occurs around the LCST of about 33 °C. This temperature-responsive nature of pNIPAM particles adds considerable richness to the phase behavior of suspensions of such particles which has been exploited in numerous studies[46,47]. In our system, as the temperature increases above room temperature, the decreasing particle size—and concomitant decreasing volume fraction of pNIPAM particles—causes the depletion-induced attraction between particles to weaken[48]. This weakening of the attraction can lead a sample which has phase separated at room temperature to become mixed and homogenous at higher temperatures (see Fig. S1 and Movie S2). However, as the temperature goes above the LCST, the pNIPAM particles will stick together due to their increased hydrophobicity. If colloid-rich droplets remain at temperatures above the LCST (by, for example, heating the sample sufficiently fast), pNIPAM particles in those droplets will aggregate into regions of colloidal gels.

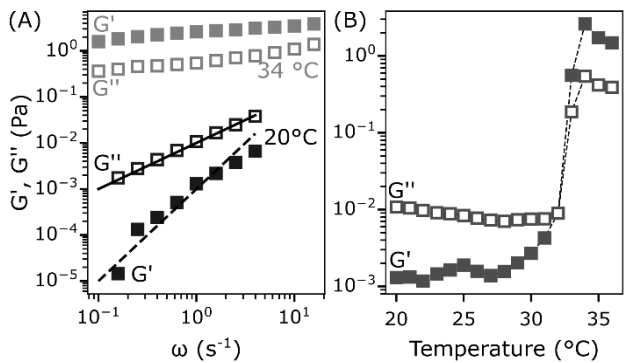

Figure 3 Rheology of the colloid-rich phase reveals the transition from a fluid-like state below the LCST to a gel-like state above. (A) Oscillatory measurements performed with a strain amplitude of 3% show liquid-like behaviour at 20 °C and gel-like behaviour at 34 °C. The storage (G′) and loss (G″) modulus are plotted as a function of the oscillation frequency. At the lower temperature, G″ > G′ while at the higher temperature G′ > G″. (B) The storage and loss moduli at the frequency of 1 rad/s are shown as a function of temperature. A sharp transition is observed at about 33 °C. Below this temperature, the sample is liquid-like while above 33 °C the sample has gelled.

The gel-like nature of the colloid-rich phase above the LCST is demonstrated through rheology. We measure the loss modulus, $G''$, and the storage modulus, $G'$, as a function of both oscillation frequency and temperature as shown in Fig. 3. At 20 °C, we observe that $G'' > G'$ and that $G' \sim \omega^2$ and $G'' \sim \omega$, indicative of a colloidal fluid. Just above the LCST, we see that $G' > G''$ and that both moduli exhibit only a weak dependency on the frequency. This is the expected behaviour of a colloidal gel. Looking at the moduli as a function of temperature, we see the storage modulus increase by three orders of magnitude and become greater than the loss modulus at roughly 33 °C.

To create microcapsules, we start with a pNIPAM-xanthan mixture which has phase separated at room temperature such that colloid-rich droplets exist within a colloid-poor continuous phase. By using a heating gun, we bring the sample's temperature above the LCST in approximately 10-20 seconds by holding the heating gun ~20 cm from the sample. Slower heating of the sample may cause the sample to mix and become homogenous before reaching the LCST because, as previously described, the depletion attraction weakens as the particles shrink. Therefore, a fast heating of the sample preserves colloid-rich droplets as the sample's temperature crosses the LCST. The use of the *rate* of temperature increase to obtain different states of mixtures containing pNIPAM particles has also been used in prior studies[49, 50].

Above the LCST, particles in the colloid-rich droplets will aggregate into a rigid cluster. This transition from a colloid-rich liquid droplet to a gelled aggregate is clearly observed under the microscope. The refractive index difference between the swollen pNIPAM particles and the suspending medium is much less than that between the shrunken PNIPAM and the suspending medium. Therefore, we see much greater contrast in the DIC micrographs above the LCST (e.g., the images at t = 0 s and at t = 50.9 s in Fig. 1B).

After seeing colloid-rich liquid droplets form spheres of gelled particles, we let the sample passively cool back to room temperature. One might expect, as we initially had, that gelled droplets would simply reliquefy as the temperature dropped below the LCST. That is, our initial expectation was that a movie of our sample being cooled from above the LCST to room temperature would look just like a movie of the sample being heated from room temperature to above the LCST run in reverse. Interestingly, that is not what we observe. Instead, we observe that as the gelled droplets cool, voids within the droplets appear. Once droplets reliquefy, a number of them contain inner droplets (see Figs. 1B, S2, S3 and Movies S1, S3, S4).

Comparison of our observations to prior work on creating double or complex emulsions help to explain this phenomenon. A general principle that has been used numerous times to create complex emulsions is phase separation within the droplet phase[51-56]. This can be achieved using solvent extraction where droplets of a mixed and homogenous fluid contain some solvent which leaves the droplet over time. What remains inside the droplet becomes more concentrated and this can lead to phase separation within the droplet, giving rise to a multiple emulsion. Phase separation can also be temperature induced[57]. The fluids within droplets may be miscible at certain temperatures and then brought to a different temperature where phase separation occurs. In our samples, as the colloid-rich droplets cool from their gelled state we observe numerous droplets appearing within which often coalesce into a single inner droplet. This seems consistent with the idea of phase separation occurring. However, it raises the question of why phase separation occurs.

We propose that the concentrations of pNIPAM and xanthan within the colloid-rich droplets are, just as the droplet reliquefies, in the unstable region of the phase diagram. While the concentrations of pNIPAM and xanthan within the colloid-rich droplets at room temperature would be placed directly on the binodal line, this binodal line shifts with temperature. At the moment that the droplet begins reliquefying, and so is just below the LCST, the binodal line has moved, relative to its room-temperature location, towards lower concentrations. This is depicted graphically in Fig. 4. Therefore, the colloid-rich droplet is unstable and separates into colloid-rich(er) and colloid-poor regions. The colloid-poor regions coalesce and lead to a single inner droplet surrounded by a shell of the colloid-rich phase.

To test this idea, we create a sample of pNIPAM and xanthan that lies just outside the room-temperature binodal line. That is, the sample is well mixed and homogenous at room temperature. If at a higher temperature the binodal line shifts downwards enough to place this sample's concentration within the unstable region of the phase diagram, then we would expect to see fluid-fluid phase separation at this higher temperature. That is what we observe. We create a sample with $\phi_p$ = 11.4% and xanthan concentration of 0.04 wt%. This sample is homogeneous when observed using a 20× objective on our microscope at room temperature. We slowly heat the sample using a Tokai Hit stage top incubator. Once the temperature is above the LCST, we observe the emergence of pNIPAM particle clusters. This observation is consistent with pNIPAM particles having attractive interparticle interactions above the LCST. We then allow the sample to cool. Rather than observing the particle clusters disappear and the solution returning to a homogeneous state, we instead see that the particle clusters form colloid-rich liquid droplets (see Fig. S4 and Movie S5). As the sample cools further towards room temperature, the sample does become homogeneous.

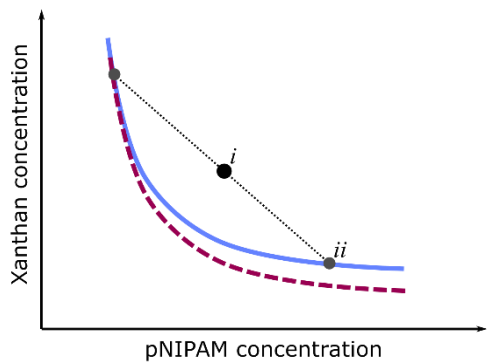

Figure 4 Cartoon aiding the explanation of core-shell droplet formation. Suppose a sample is made composition placing it at point (i). As it is within the unstable region of the phase diagram, phase separation will occur with the colloid-rich phase having a composition corresponding to point (ii). If the binodal line shifts from the blue (the room temperature binodal) to the dashed red (the higher temperature binodal) line, then the point at (ii) becomes unstable. The droplet having a composition corresponding to point (ii) will then undergo phase separation.

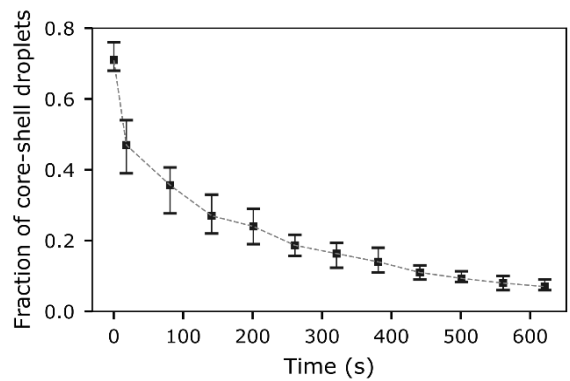

Figure 5 Over the course of several minutes, most inner droplets escape. We monitor a volume of 290×269×100 μm³ over the course of several minutes immediately after forming core-shell droplets. Initially, about 70% of the 1176 droplets within this region contain inner droplets. That fraction falls to about 7.5% in 10 minutes. To generate the error bars, we split the monitored volume into three and show the minimum and maximum fractions of core-shell droplets among the three regions.

We note that our observation of liquid-liquid phase separation occurs only as the sample cools and not as it is heated. We speculate that since heating above the LCST leads to clusters of particles, those clusters then form colloid-rich fluid droplets as the sample cools more easily than fluid droplets nucleate from the homogeneous state as the sample heats. In future work, we plan to more controllably and slowly varying the sample temperature. Perhaps if we could maintain the temperature of the sample below the LCST for an extended period of time, we would eventually see fluid colloid-rich droplets nucleate and grow.

We have demonstrated how a mixture of pNIPAM colloidal particles and xanthan can be made to form a double emulsion after a ~15 °C increase and then decrease in temperature. However, the core-shell droplets that form are not long lasting. As the shells consist of a colloid-rich phase that is fluid, the inner droplets escape and join the continuous colloid-poor phase after a few minutes. In Fig. 5, we show the fraction of droplets which contain an inner droplet as a function of time. Immediately after the heating and cooling of our sample, we observe that ~70% of 1176 droplets within a volume of 290×269×100 μm³ contain inner droplets. After a couple minutes, less than a third of all droplets have an inner droplet and after 10 minutes, less than 10% do. The typical time it takes for inner droplets to escape in double emulsions has been explored theoretically by Datta et al.[58] and, according to their model, it would be possible to extend the inner droplet lifetime by altering the density difference between the two phases, the viscosity of the shell, and/or the size of the inner and outer droplets. We do note that the core-shell droplets which are the longest lasting tend to be on the higher end of the size distribution of droplets (see Fig. S5).

We chose to make longer lasting microcapsules with a solid shell by taking advantage of pNIPAM particles' response to salt. At higher salt concentrations, the interaction between pNIPAM particles is increasingly attractive and pNIPAM particles will aggregate. This is due to salt screening the electrostatic repulsion between particles and has been noted in numerous studies[38,59-61]. Therefore, by increasing the salt concentration immediately after the double emulsion forms, the colloid-rich fluid shell should form a solid gel. We devise a simple strategy to confirm that this method for producing microcapsules with a solid shell works. Out of a glass slide and coverslip we make sample holders as usual for observation on our optical microscope. However, we add a few grains of NaCl between the slide and coverslip to one side of the sample holder. After pipetting in our mixture of pNIPAM particles and xanthan, the NaCl will slowly dissolve and spread throughout the sample. By immediately heating the sample to a temperature above the LCST with the heat gun, we observe core-shell droplets form in regions of the sample away from the NaCl grains. As the NaCl dissolves and diffuses through the sample, we observe that the core-shell droplets form capsules with rigid shells. An image of a microcapsule with a solid shell of pNIPAM particles is shown in Fig. 6 and Movie S6 captures the process.

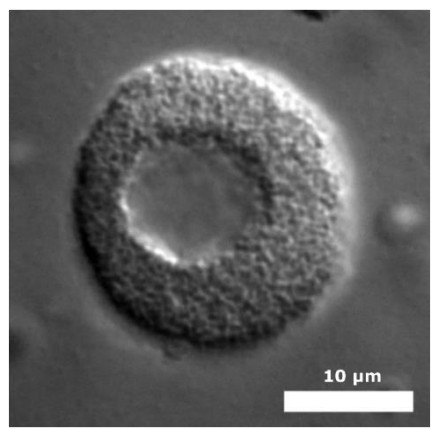

Figure 6 Microcapsule with a solid shell at room temperature. After forming core-shell droplets through raising the temperature above the LCST of pNIPAM and then allowing the sample to cool, we increase the NaCl concentration such that the pNIPAM particles aggregate and form a solid shell.

## 4 Conclusions

We fabricate core-shell droplets and microcapsules with rigid shells from a colloid-polymer mixture consisting of temperature-sensitive pNIPAM microgel particles and xanthan. At room temperature the colloid-polymer mixture phase separates into colloid-rich droplets in a continuous colloid-poor phase. An increase in temperature above the LCST of pNIPAM results in an aggregation of the pNIPAM particles and the gelation of the colloid-rich fluid droplets. As the sample cools, the colloid-rich regions reliquefy and undergo a fluid-fluid phase separation which leads to colloid-poor inner droplets. These all-aqueous core-shell droplets typically last no more than a few minutes before the inner droplet escapes. However, an increase in salt concentration will cause the pNIPAM particles to aggregate and cause the colloid-rich shell to solidify.

Our presented method uses all aqueous components making it suitable for biological material that would be damaged by organic solvents. Additionally, the use of colloidal particles to form the shell imparts permeability to the shell which could be tuned by adjusting the size of the colloidal particles. Finally, microcapsules that are responsive to multiple stimuli may find a range of applications[62,63]. With shells of pNIPAM microgel particles, one could take advantage of these particles' responsiveness to changes in temperature, pH and ionic strength to have the microcapsules rupture and release their content when exposed to certain stimuli.

For certain encapsulation applications, capsules of uniform size are ideal. Our method starts with an emulsion of colloid-rich droplets created by manually shaking which produces a very polydisperse collection of droplets. Future work could employ microfluidics to create monodisperse capsules[64]. A further advantage of using microfluidics would be that heating of the core-shell droplets above the LCST and an increase in salt concentration could all be done on-chip. Finally, we note that the mechanical strength and permeability of microcapsule shells are important parameters which we have yet to explore for our system. However, we hope that the method we have described, which uses a colloid-polymer system with water as the common solvent, finds use in applications where stimuli-responsive all-aqueous microcapsules are required.

## Conflicts of interest

There are no conflicts to declare.

## Acknowledgements

We acknowledge support by a Cottrell Scholar Award to RM from the Research Corporation for Science Advancement (RCSA), a Cottrell Instrumentation Supplements Award from RCSA (Award #27459), and support by NSF (CBET-1919429).

# Electronic Supplementary Information for
# Core-shell droplets and microcapsules formed through liquid-liquid phase separation of a colloid-polymer mixture

Steven Dang, John Brady, Ryle Rel, Sreenidhi Surineni, Conor O'Shaughnessy, and Ryan McGorty

*Department of Physics and Biophysics, University of San Diego, San Diego, CA 92110 USA*

**This PDF file includes:**
   Figures S1 – S5
   Captions to Movies S1 – S6

**Other Supplementary Materials include the following:**
   Movies S1 – S6

**Location of Movies:**
https://drive.google.com/drive/folders/1etOw7mbBaPGRhxLWXWQINJh8zBteG4_A?usp=sharing

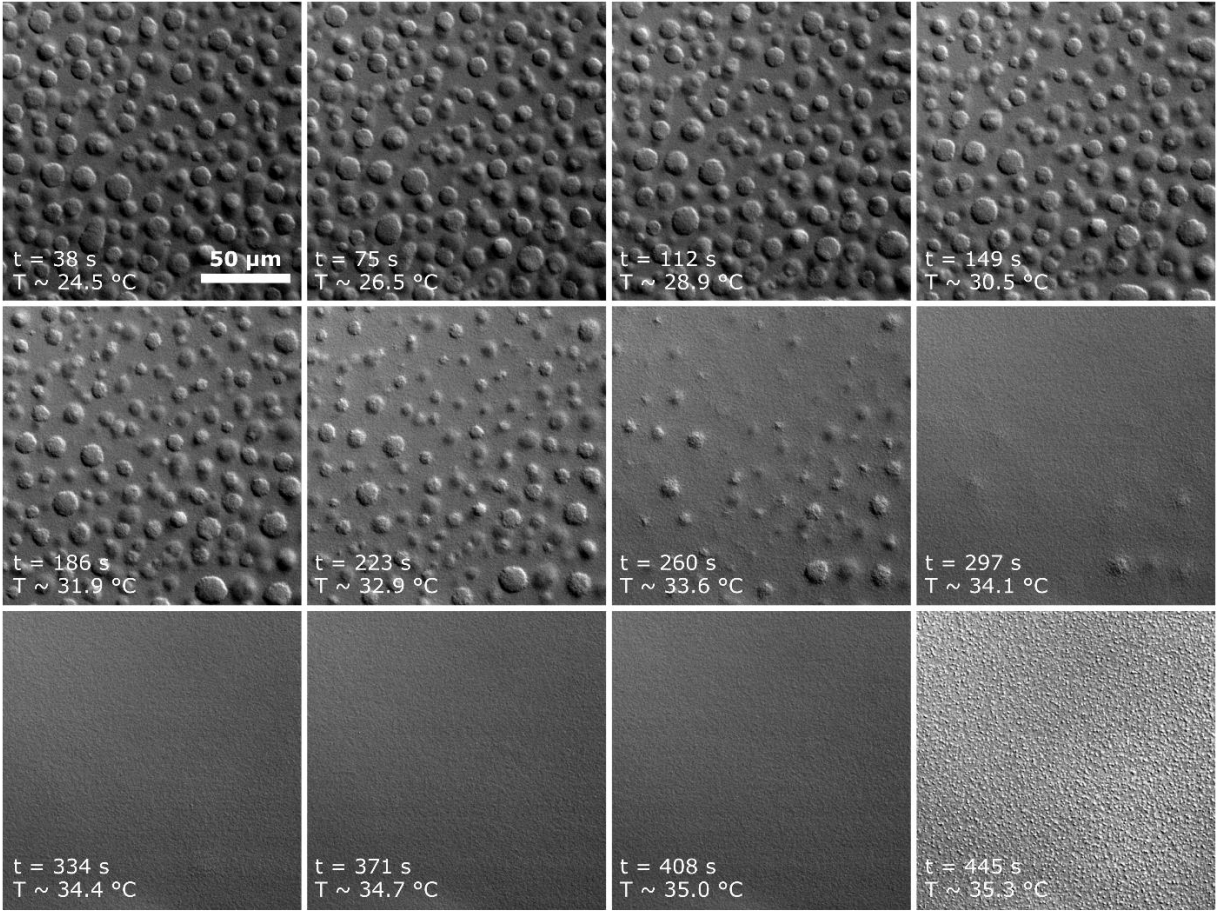

**Fig. S1.** A gradual increase in temperature results in the phase separated colloid-polymer mixture becoming homogeneous. A pNIPAM-xanthan mixture (0.08 wt% xanthan, 11.4% pNIPAM volume fraction) separates into colloid-rich droplets in a continuous colloid-poor phase at room temperature. Using a stage-top incubator on an inverted light microscope we heat the sample above the LCST of pNIPAM. As the temperature increases, the colloid-rich droplets shrink and disappear until the sample becomes homogeneous. Above the LCST, we see clusters of pNIPAM particles. The temperature indicated in the figure is recorded from a temperature probe placed a couple of centimeters away from the sample but on the glass slide. However, since the objective lens (a 20× air objective) was not heated, we suspect the actual sample temperature to be approximately 1-2 °C lower. See Movie S2.

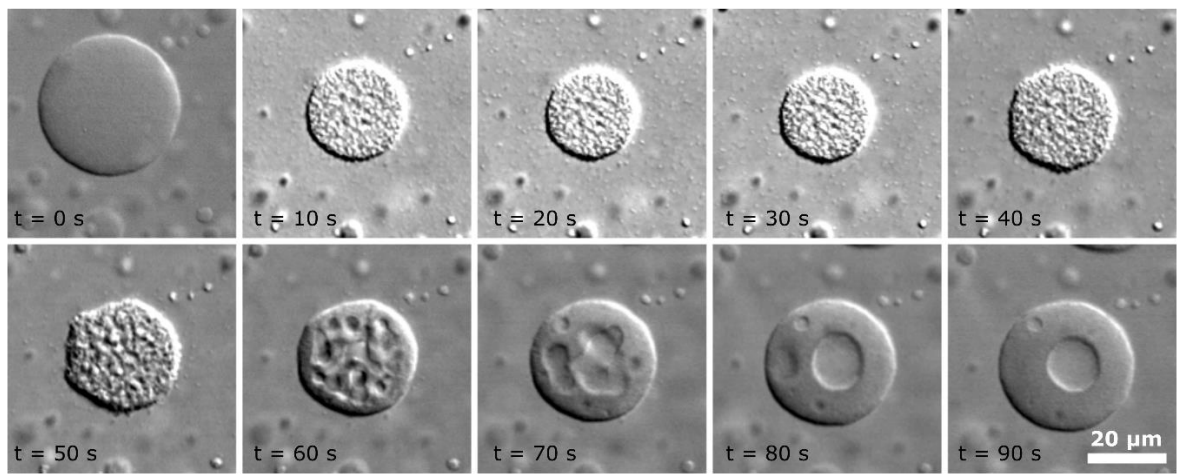

**Fig. S2.** Formation of a core-shell droplet occurs after a single emulsion droplet is heated and cooled. A droplet of the colloid-rich phase within a colloid-poor phase is shown at t = 0 s. The sample is heated for approximately 10 seconds using a heat gun. The temperature goes above the LCST of pNIPAM since the pNIPAM particles shrink and aggregate. As the sample passively cools, the colloid-rich droplet reliquefies, and numerous colloid-poor droplets can be seen inside (t=60s). As those internal droplets coalesce, a core-shell droplet forms. See Movie S3.

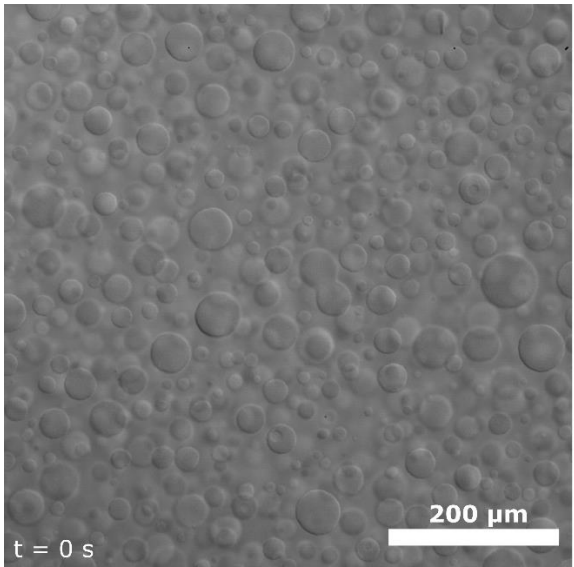
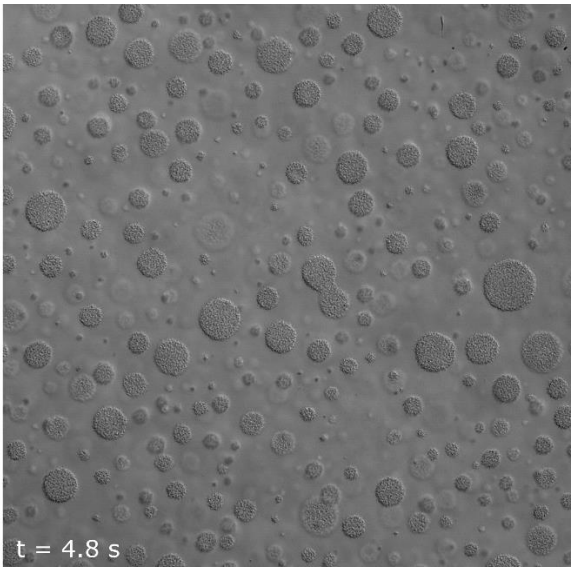
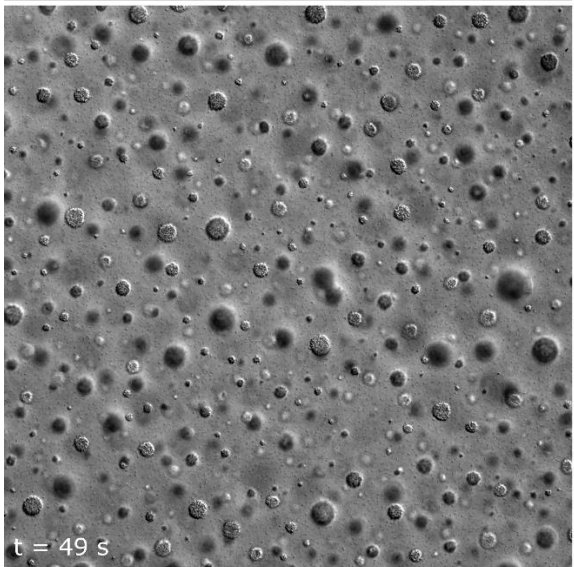
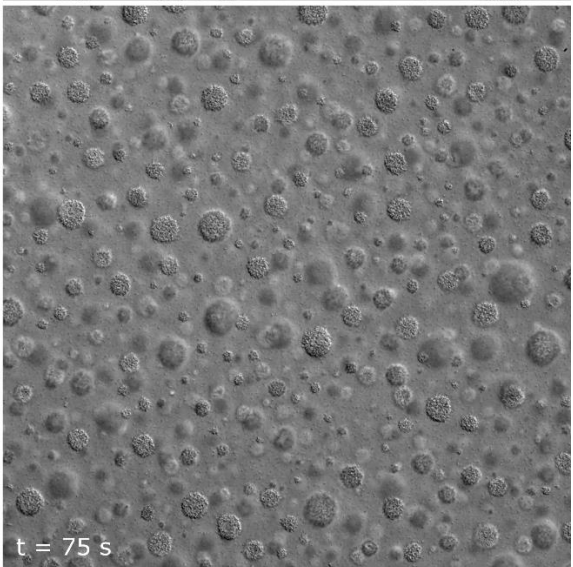
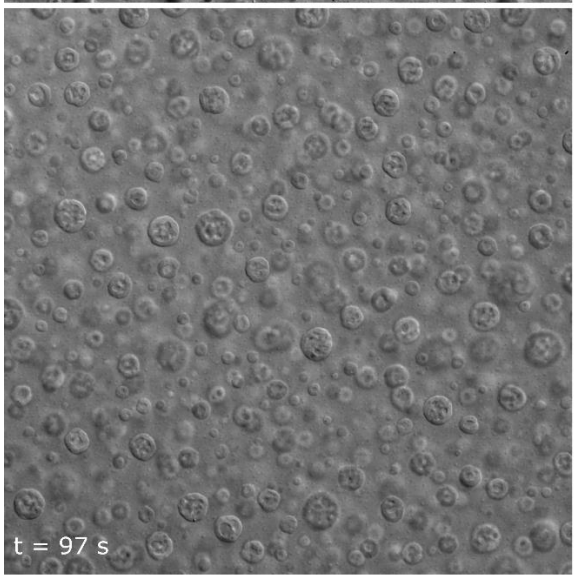
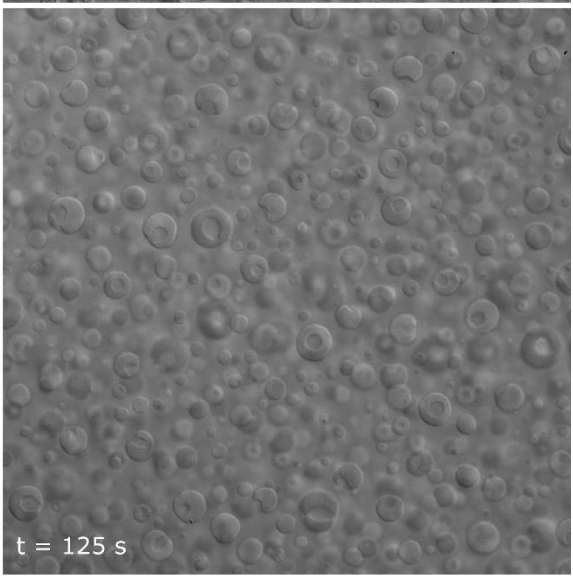

**Fig. S3.** Formation of core-shell droplets occurs after the sample is heated and cooled. Images show the same process as in Fig. S2 over a wider field of view. See Movie S4.

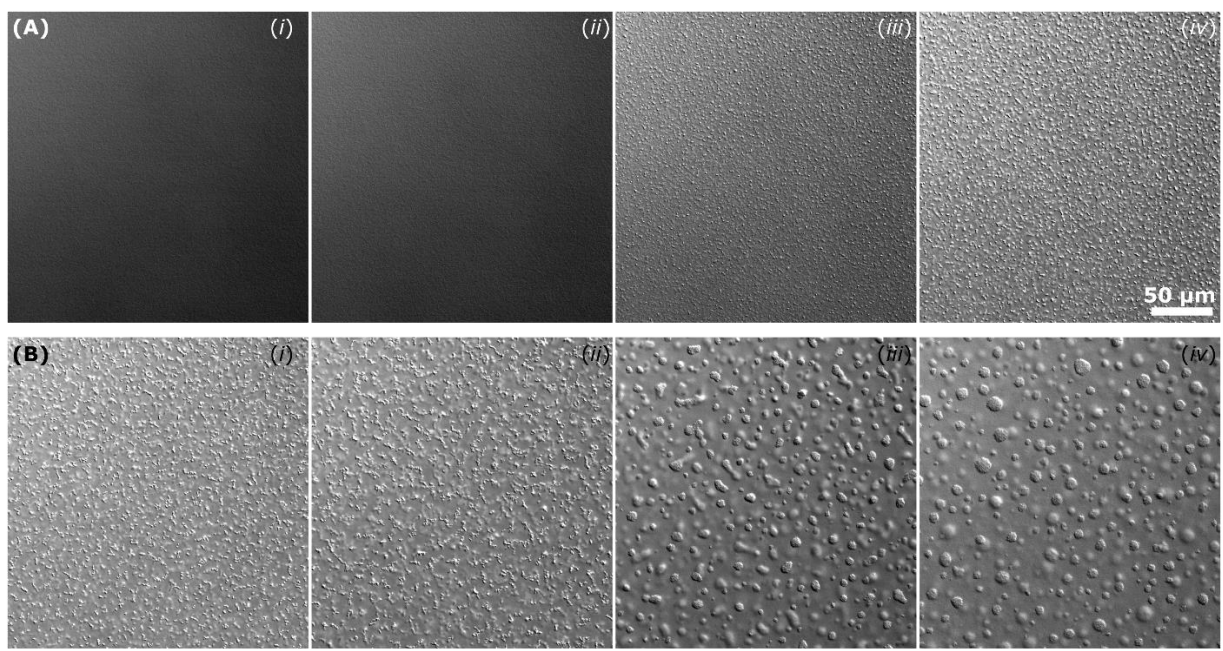

**Fig. S4.** A pNIPAM-xanthan mixture which is homogeneous at room temperature shows liquid-liquid phase separation at just below the LCST as it cools from above the LCST. (A) A sample that is mixed at room temperature (0.04 wt% xanthan, 11.4%) is slowly heated using a stage top incubator on an inverted light microscope. The frames shown as (*i*) through (*iv*) are each separated by 30 s and the temperature measured near the sample over this entire time span goes from approximately 34 to 34.2 °C. Clusters of pNIPAM particles are observed in (*iii*) and (*iv*). (B) Following the slow heating in (A), the sample is then cooled. The frames shown as (*i*) through (*iv*) are each separated by 120 s and the temperature measured near the sample over this entire time span goes from approximately 34.2 to 33.2 °C. We observe the pNIPAM particle clusters forming colloid-rich fluid droplets. As the sample cools back to room temperature, these droplets disappear, and the sample becomes homogeneous. We note that the temperature of the sample is slightly less than that measured by the probe as explained in the caption to Fig. S1. See Movie S5.

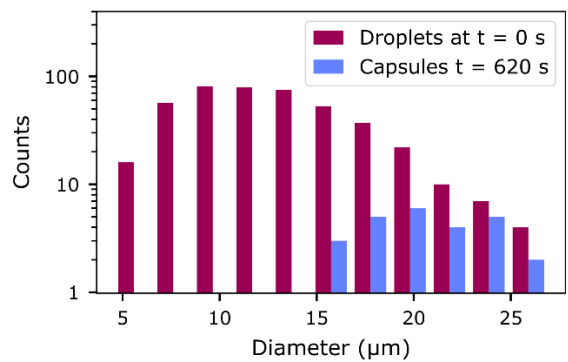

**Fig. S5.** The size distribution of droplets within a volume of 290×269×100 µm³ is shown at room temperature before the sample is heated (at t = 0 s) in red. About 10 minutes after heating the sample to generate capsules, about 7.5% of droplets contain an inner droplet (as shown in Fig. 5). The size distribution of those capsules which remain is shown in blue.

**Caption to Movie S1.** The process of core-shell droplet formation is shown. This movie corresponds to the images shown in Fig. 1B. The sample (0.08 wt% xanthan, 11.4% volume fraction pNIPAM) was heated with a heat gun for approximately 20 seconds. As the sample cools, phase separation within the droplet is seen which results in a single colloid-poor droplet within the colloid-rich droplet. The scale bar is 25 µm. Playback speed is 10× real speed.

**Caption to Movie S2.** A slow temperature increase causes a demixed to mixed transition. This movie shows a sample with the same composition as the sample imaged in Fig. 1 and Movie S1 (0.08 wt% xanthan, 11.4% volume fraction pNIPAM). However, unlike in Movie S1, the sample is slowly heated using a stage top incubator. Individual frames from this movie are shown in Fig. S1. The scale bar is 50 µm. Playback speed is 100× real speed.

**Caption to Movie S3.** The process of core-shell droplet formation is shown. This movie corresponds to the images shown in Fig. S2. The scale bar is 20 µm. Playback speed is 2× real speed.

**Caption to Movie S4.** The process of core-shell droplet formation is shown. This movie corresponds to the images shown in Fig. S3. The scale bar is 200 µm. Playback speed is 10× real speed.

**Caption to Movie S5.** Fluid-fluid phase separation observed after a decrease in temperature from above the LCST in a sample homogeneous at room temperature. This movie corresponds to the images shown in Fig. S4. The scale bar is 50 µm. Playback speed is 100× real speed.

**Caption to Movie S6.** Solidification of the colloid-rich shell. This movie corresponds to the image shown in Fig. 6. A core-shell droplet was formed in the same way as shown in Movies S1, S3, and S4. The concentration of NaCl in the neighborhood of the droplet then increased which caused the colloid-rich shell to solidify. The colloid-rich shell is fluid in the beginning of this movie, but is rigid by the end. The scale bar is 10 µm. Playback speed is 4× real speed.